\documentclass[epj]{svjour}

\def\makeheadbox{\relax}

\usepackage[T1]{fontenc}
\usepackage[utf8]{inputenc}
\usepackage{amsmath}
\usepackage{amssymb}
\usepackage{graphicx}

\newcommand{\diffd}{\mathrm{d}}
\newcommand{\kb}{k_{\mathrm{B}}}
\newcommand{\piint}{\int\limits_0^{2\pi}}
\newcommand{\refeq}[1]{Eq.\ (\ref{#1})}
\newcommand{\reffig}[1]{Fig.\ \ref{#1}}
\newcommand{\refsec}[1]{Sec.\ \ref{#1}}

\begin{document}

\title{Kramers escape of a self-propelled particle}

\author{Alexander Geiseler \and Peter H\"anggi \and Gerhard Schmid}

\institute{Institute of Physics, University of Augsburg, Germany}

\date{Submitted: \today}

\abstract{
We investigate the escape rate of an overdamped, self-propelled spherical Brownian particle on a surface from a metastable potential well. Within a modeling in terms of a 1D constant speed of the particle's active dynamics we consider the associated rate using both numerical and analytical approaches.
Regarding the properties of the stationary state in the potential well, two major timescales exist, each governing the translational and the rotational dynamics of the particle, respectively. The particle radius is identified to present the essential quantity in charge of regulating the ratio between those timescales. For very small and very large particle radii, approximate analytic expressions for the particle's escape rate can be derived, which, within their respective range of validity, compare favorably with the precise escape numerics of the underlying full two-dimensional Fokker-Planck description.
\PACS{
      {05.40.-a}{Fluctuation phenomena, random processes, noise, and Brownian motion}   \and
      {05.10.Gg}{Stochastic analysis methods (Fokker-Planck, Langevin, etc.)}
     }
}
\maketitle

\section{Introduction}
\label{SecIntro}

The dynamics of self-propelled Brownian particles (SPPs) increasingly attracts the attention of researchers in recent years \cite{Romanczuk2012,Walther2013,Elgeti2015Rev}. As opposed to conventional Brownian motion, where the dynamics of a particle is determined solely by the movement of the surrounding gas or liquid
molecules, here the particle possesses in addition an internal propulsion mechanism. In principle, this propulsion is generated either by a local non-equilibrium in the vicinity of the particle, which can be, for example, \ of thermo- \cite{Wurger2007,Jiang2010,Buttinoni2012,Yang2013} electro- \cite{Moran2010,Ebbens2014} or diffusiophoretic \cite{Golestanian2005,Howse2007,Volpe2011,Drube2013} nature, or by an active deformation of the particle's shape, leading to a ``swimming'' behavior \cite{Lighthill1976,Lauga2009}. The diverse properties and effects that are inherent in this self-propulsion provide a large resource for applications for artificial SPPs (which are also called Janus particles), such as nano-robots and drug carriers \cite{Paxton2006,Sundararajan2008,Baraban2012}. In addition, many biological processes can be described well using a self-propulsion model, e.g.\ the movement of the bacteria \emph{Myxococcus xanthus} and \emph{Escherichia coli} \cite{BergEColi,Gejji2012,Elgeti2015}.

In the following, we study the escape rate $\mathrm{\Gamma}$ of a spherical SPP on a surface out of a metastable potential well. The underlying fundamental problem of a Brownian particle's escape over a potential barrier is known as the Kramers escape problem, named after H. A. Kramers \cite{Kramers1940}.
Since Kramers’ pioneering publication, the field of escape dynamics has been generalized and advanced considerably, including both quantum escape and various non-equi\-lib\-ri\-um settings---a comprehensive overview of the state-of-the-art is provided with Refs. \cite{Hanggi1986,hanggiAddendum1986,Hanggi1990}.

Up to now, only a limited number of works exist on the objective of escape dynamics of SPPs in different settings \cite{Pohlmann1997,Burada2012,Ghosh2014,Schaar2015}. In the item \cite {Burada2012}, which as well addresses the escape from a metastable potential, the authors used a non-linear friction coefficient to model the particle's propulsion; this approach thus distinctly differs from the modeling here, involving the role of rotational Brownian motion (see \refsec{SecModel}). This in turn renders the full escape dynamics more complex, involving a Fokker-Planck description for both the planar position and the rotational angle degrees of freedom.

In order to obtain analytic results for the aforementioned escape rate, we will analytically study two limiting cases, namely the case of a slow rotation dynamics and the case of a very fast particle rotation.

\section{Model Setup}
\label{SecModel}

\begin{figure}[t]
\begin{center}\includegraphics[width=0.55\columnwidth]{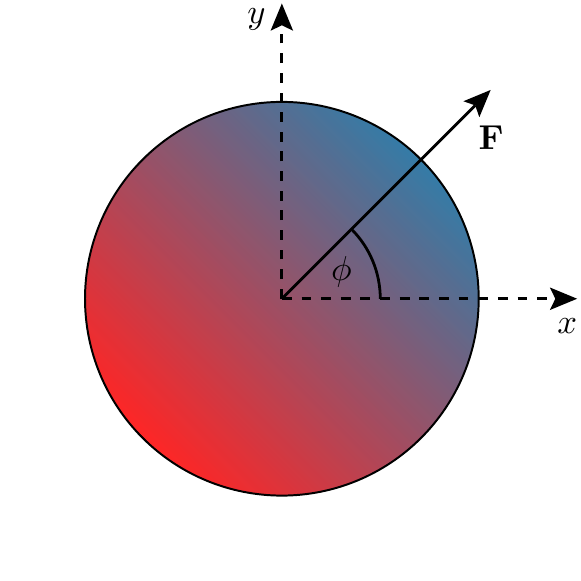}\end{center}%
\caption{\label{F1}(color online) Sketch of our model of a two-dimensional SPP. The propulsion force $\mathbf{F}$---which is assumed to be of constant magnitude---acts along a specific direction of the particle's orientation.}
\end{figure}

To mathematically model the self-propelled, spherical particle's dynamics on a surface occurring in a metastable potential landscape, we start out from a 2D over-damped Brownian particle in an external metastable potential $U(\mathbf{r})$ that in addition to the thermal fluctuations is driven by a self-propulsion force $\mathbf{F}$. The external potential $U(\mathbf{r})$ can experimentally be realized by use of two scanned laser tweezers, as demonstrated \emph{in situ} for the phenomenon of stochastic resonance and resonance activation \cite{Schmitt2006}. Furthermore, it is important to stress that the force $\mathbf{F}$---which is caused by one of the propulsion mechanisms mentioned in \refsec{SecIntro}---is not external, but rather is inherent to the particle. The propulsion acts along a specific direction $\mathbf{n}$ of the particle's orientation (see \reffig{F1}), with the latter also subjected to rotational fluctuations. Consequently we can model the escape dynamics with a multi-dimensional Langevin dynamics of the form
\begin{alignat}{2}
\label{Langevin1}
&\diffd \mathbf{r}&&=\frac{D}{\kb T}\left[F\mathbf{n}-\nabla U(\mathbf{r})\right]\diffd t+\sqrt{2D}\,\diffd \mathbf{W}_\mathrm{t}\\
\label{Langevin2}
&\mathbf{n}&&=\left(\begin{array}{r}\cos\phi\\ \sin\phi\end{array}\right)\\
\label{Langevin3}
&\diffd\phi&&=\sqrt{2D_\mathrm{r}}\,\diffd W_\mathrm{r}\;,
\end{alignat}
where $\mathbf{n}$ is parameterized by the angle $\phi$ that also performs a rotational Brownian motion. Here, $D$ and $D_\mathrm{r}$ denote the translational and rotational diffusion constant, respectively, $T$ characterizes the temperature of the surrounding fluid, $\kb$ is Boltzmann's constant and $\nabla$ the Cartesian gradient. The stochastic processes $\mathbf{W}_\mathrm{t}$ and $W_\mathrm{r}$ are standard Wiener processes with mean zero and variance $t$, i.e., corresponding to Gaussian white noise of unit strength.

For the case of a spherical particle and in presence of a low Reynolds number dynamics, $D_\mathrm{r}$ can be expressed by $D$ according to $D_\mathrm{r}=3D/(4R^2)$, with $R$ being the particle's radius \cite{Serdyuk2007}. The Fokker-Planck equation associated with the Langevin dynamics, Eqs.\ (\ref{Langevin1}), (\ref{Langevin2}) and (\ref{Langevin3}), thus reads
\begin{alignat}{2}
\frac{\partial P(\mathbf{r},\phi,t)}{\partial t}=&D\left[\Delta+\nabla\left(\frac{\left[\nabla U(\mathbf{r})\right]}{\kb T}-\frac{F}{\kb T}\mathbf{n}\right)\right.\nonumber\\
&\quad\ \,\left.+\frac{3}{4R^2}\frac{\partial^2}{\partial\phi^2}\right]P(\mathbf{r},\phi,t)\;,
\end{alignat}
with $\Delta$ denoting the Cartesian Laplace operator. Assuming that $U(\mathbf{r})$ depends on the coordinate $x$ only, it is convenient to focus only on the dynamics of the $x$ component of the particle's position. This is legitimate since the $y$ coordinate may be integrated out from the latter equation, leading to an equation for the marginal probability density $P(x,\phi,t)$,
\begin{alignat}{2}
\label{FPE1D}
\frac{\partial P(x,\phi,t)}{\partial t}=&D\left[\frac{\partial}{\partial x}\left(\frac{\partial}{\partial x}+\frac{U'(x)}{\kb T}-\frac{F}{\kb T}\cos\phi\right)\right.\nonumber\\
&\quad\ \,\left.+\frac{3}{4R^2}\frac{\partial^2}{\partial\phi^2}\right]P(x,\phi,t)\;,
\end{alignat}
where the prime denotes the derivative w.r.t.\ $x$.

\section{Kramers Rate for Self-Propelled Particles}
\label{SecKramers}

\begin{figure}[t]
\begin{center}\includegraphics[width=0.8\columnwidth]{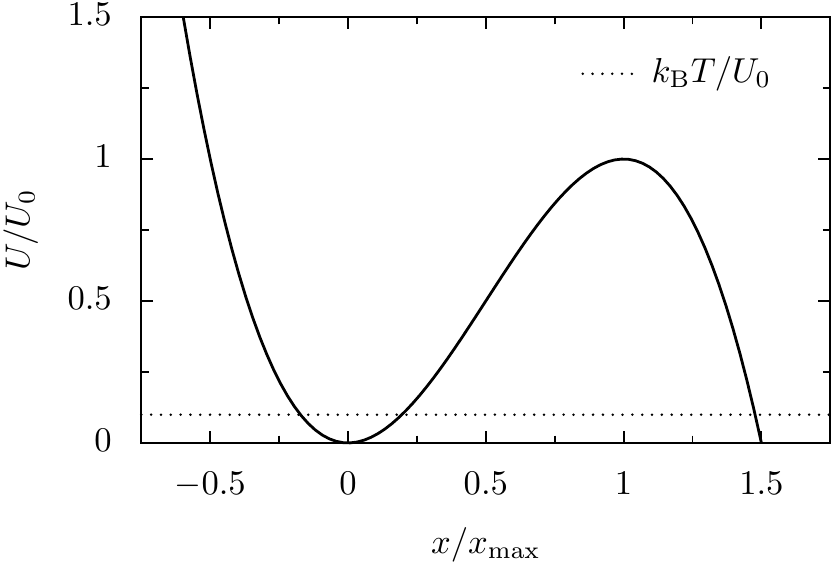}\end{center}%
\caption{\label{F2}Plot of the metastable potential $U(x)$ as considered in the present paper for $k=10^{-6}\,\mathrm{N}\mathrm{m}^{-1}$ and $x_\mathrm{max}=0.5\,\mu m$. It can clearly bee seen that for this choice of parameters the barrier is much higher than the thermal energy (indicated by the dotted line; $T=300\,\mathrm{K}$), why for passive particles (i.e., for $F=0$) escapes are very rare events.}
\end{figure}

To start with, we proceed from the Fokker-Planck equation (\ref{FPE1D}), where for concreteness the metastable potential $U(x)$ is next assumed to take on a cubic shape, reading explicitly $U(x)=(1/2)kx^2[1-2x/(3x_\mathrm{max})]$ (see \reffig{F2}). The only restriction on this rather general cubic potential is that the spring constant of the potential's harmonic approximation at the bottom of the well and on top of the barrier (namely $k$ and $-k$) possess the same absolute value. However, we remark that the generalization to the case of two different hook constants is straightforward and can readily be implemented, using the same methods as employed in the present work.

Introducing a spatial and temporal dimensionless scaling, i.e., $x=:x_\mathrm{max}\xi$ and $t=:\kb T/(Dk)\tau$, where $\kb T/(Dk)\allowbreak=:t_k$ is the relaxation time of the particle (due to the restoring force of the potential) for small deflections from its equilibrium position at $x=0$, we end up with the dimensionless, two-dimensional Fokker-Planck equation
\begin{alignat}{2}
\label{FPE1Dscaled}
\frac{\partial P(\xi,\phi,\tau)}{\partial \tau}=&\left[\frac{\partial}{\partial \xi}\left(\frac{\kb T}{6U_0}\frac{\partial}{\partial \xi}+\xi(1-\xi)-\frac{F}{kx_\mathrm{max}}\cos\phi\right)\right.\nonumber\\
&\quad\ \,\left.+\frac{3\kb T}{4kR^2}\frac{\partial^2}{\partial\phi^2}\right]P(\xi,\phi,\tau)\;.
\end{alignat}
Here, the factor $\kb T/(6U_0)$ is proportional to the ratio between the thermal energy and the barrier height $U_0:=U(x_\mathrm{max})-U(0)=(1/6)kx_\mathrm{max}^2$. Furthermore, $F/(kx_\mathrm{max})$ characterizes the ratio between the propulsion force and the restoring force of the potential and $3\kb T/(4kR^2)$ denotes the ratio between $t_k$ and the rotational diffusion time constant $t_\mathrm{r}=D_\mathrm{r}^{-1}$ \cite{Zheng2013}. The escape dynamics of this two-dimensional Fokker-Planck dynamics is detailed within Appendix \ref{NumMeth}.

\subsection{Fixed Angle Approximation}

\begin{figure}[t]
\begin{center}\includegraphics[width=0.8\columnwidth]{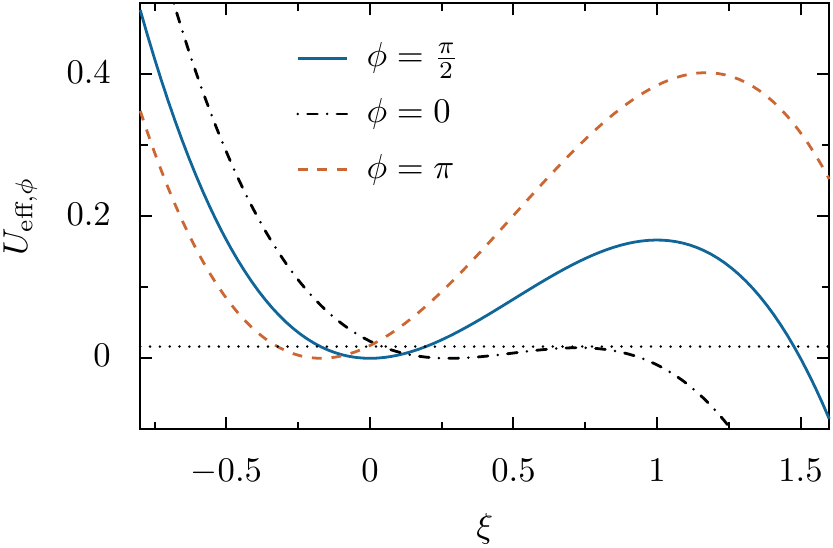}\end{center}%
\caption{\label{F3}(color online) Plot of the tilted effective potential $U_{\mathrm{eff},\phi}(\xi)$ for different orientations of the SPP, where the choice of parameters was $F=10^{-13}\,\mathrm{N}$, $k=10^{-6}\,\mathrm{N}\mathrm{m}^{-1}$, and $x_\mathrm{max}=0.5\,\mu\mathrm{m}$. The curves were offset, so that the minimum value of $U_{\mathrm{eff},\phi}$ is always equal to zero. Regarding the present choice of parameters, the potential barrier for particles orientated to the left is much higher than the scaled thermal energy $\kb T/(6U_0)$ indicated by the dotted line ($T=300\,\mathrm{K}$); however, for particles orientated to the right the barrier has practically vanished.}
\end{figure}

To gain analytical insight into the particle's escape rate $\mathrm{\Gamma}$, we first concentrate on the limit of a very slow particle rotation; more specifically, on the limit $t_\mathrm{r}/t_k\to\infty$. In this limit, the particle's orientation can be regarded as fixed during an escape attempt out of the well and the partial differential equation (\ref{FPE1Dscaled}) reduces to an ordinary differential equation w.r.t.\ $\xi$ that now contains an effective, $\phi$-dependent potential, reading
\begin{equation}
\label{FPEFixedPhi}
U_{\mathrm{eff},\phi}(\xi)=\frac{\xi^2}{2}-\frac{\xi^3}{3}-\frac{F}{kx_\mathrm{max}}\,\xi\cos\phi\;.
\end{equation}
Inspecting \reffig{F3}, it becomes obvious, however, that a barrier height justifying the assumption of rare escape events for passive particles does generally not vindicate this assumption for active particles---with increasing propulsion strength, the effective barrier height becomes steadily lowered for active particles orientated to the right, until finally the barrier vanishes and the process cannot be described in terms of a rare escape anymore. The critical value of $F$ causing the local extremes of $U_{\mathrm{eff},\phi}(\xi)$ to coalesce to yield a sole saddle point is given by $F_\mathrm{crit}=kx_\mathrm{max}/(4\cos\phi)$, implying that in the fixed angle approximation $F$ is not allowed to exceed the minimal value $kx_\mathrm{max}/4$. In reality, one may expect that $F$ must be chosen at least one order of magnitude smaller.

With these limitations in mind we succeeded to reduce the escape problem of an active particle to the escape problem of a passive particle moving in a modified potential \cite{Pototsky2012}. We next can invoke the flux-over-population method, see Eqs.\ (2.26) and (2.27) in Ref.\ \cite{Hanggi1986}, to analytically calculate the escape rate. Assuming rare escape events, the particle's escape rate at fixed $\phi$, $\mathrm{\Gamma}_\phi$, can readily be obtained by calculating its stationary non-equilibrium probability current across the effective potential barrier, yielding that
\begin{alignat}{2}
\mathrm{\Gamma}_\phi=&\frac{1}{2\pi}\sqrt{1-\frac{4F\cos\phi}{kx_\mathrm{max}}}\nonumber\\
&\times\exp\left(-\frac{U_0}{\kb T}\left(1-\frac{4F\cos\phi}{kx_\mathrm{max}}\right)^\frac{3}{2}\right)\;.
\label{Gammaphi}
\end{alignat}
In order to now account for the fact that $\phi$ is not fixed, but is rather very slowly changing compared to the timescale $t_k$---i.e., with $\phi$ staying nearly fixed during an escape attempt, while undergoing thermalization on the timescale of the particle's sojourn inside the well (resulting in a uniform
angular distribution)---we are allowed to average $\mathrm{\Gamma}_\phi$ w.r.t.\ $\phi$, yielding the escape rate
\begin{equation}
\label{EscapeRateTimeSep}
\mathrm{\Gamma}=\frac{1}{2\pi}\piint\diffd\phi\,\mathrm{\Gamma}_\phi\;.
\end{equation}

\begin{figure}[t]
\begin{center}\includegraphics[width=0.85\columnwidth]{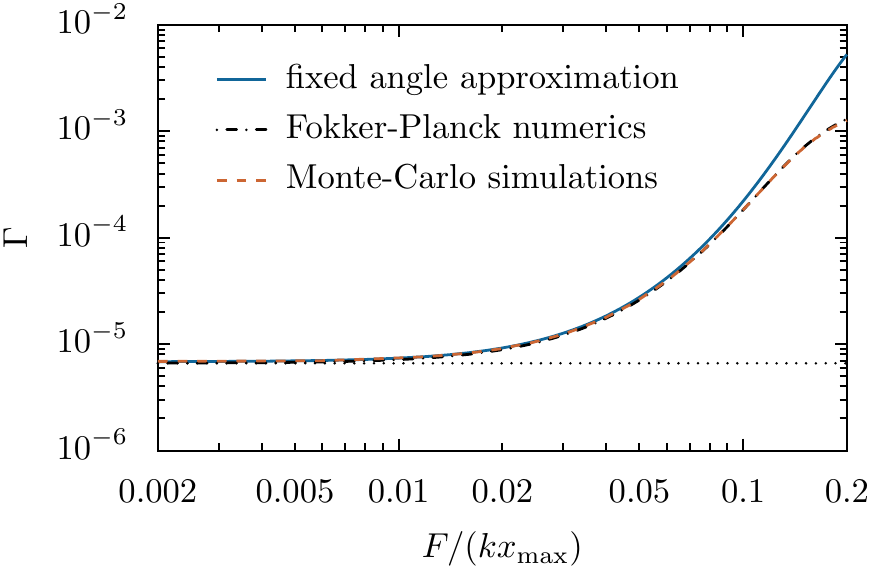}\end{center}%
\caption{\label{F4}(color online) Escape rate $\mathrm{\Gamma}$ of a SPP as a function of the propulsion strength $F$, with the parameters being $k=10^{-6}\,\mathrm{N}\mathrm{m}^{-1}$, $T=300\,\mathrm{K}$, $x_\mathrm{max}=0.5\,\mu\mathrm{m}$, and $R=1\,\mu\mathrm{m}$. For details regarding the applied numerical methods, see Appendix \ref{NumMeth}. The dotted black line indicates the exact escape rate of a passive Brownian particle. Recognizably, the fixed angle approximation yields good results for small to moderate $F$, for $F$ larger than approximately $0.1kx_\text{max}$ however, the approximation breaks down due to the fact that the particle cannot thermalize in the well anymore.}
\end{figure}%

A graphical comparison between the escape rate calculated by means of Eqs.\ (\ref{Gammaphi}, \ref{EscapeRateTimeSep}) and the precise two-di\-men\-sion\-al numerics is depicted in \reffig{F4}. The particle radius, which essentially controls the ratio $t_\mathrm{r}/t_k$ and thus the validity of the fixed angle approximation, is chosen in such a way that $t_\mathrm{r}/t_k\approx 320$, for what reason \refeq{EscapeRateTimeSep} is expected to yield applicable results. Indeed, the fixed angle approximation compares favorably with the numerical outcomes for small to moderate propulsion strengths. If the self-propulsion force $F$ however becomes too large, the approximation starts to fail, yielding unfavorable agreement. This is mainly owed to the fact that the mean escape time $\mathrm{\Gamma}_{\phi=0}^{-1}$ of particles orientated rightward becomes increasingly smaller than the scaled rotational diffusion time. For this very reason the particle's angular dynamics cannot equilibrate in the well any longer and \refeq{EscapeRateTimeSep} becomes invalid. Consequently, also for $R$ chosen too large the fixed angle approximation breaks down; this is so because $t_\mathrm{r}/t_k$ scales with $R^2$ and $\mathrm{\Gamma}_{\phi=0}^{-1}$ is independent of $R$. Hence, the fixed angle approximation has a limited range of validity regarding the size of the radius $R$: the particle must rotate so slowly that $t_\mathrm{r}\gg t_k$ (justifying the separation of rotational and translational timescales), however it also must rotate sufficiently fast so that during its sojourn in the potential well it (at least approximately) is allowed to thermalize. This yields the condition that $\mathrm{\Gamma}_{\phi=0}^{-1}\gg t_\mathrm{r}/t_k\gg 1$.

\subsection{Diffusive approximation}

In the opposite limit of fast particle rotation one again is able to obtain an analytic expression for the escape rate of a SPP; more specifically, in the limit that $t_\mathrm{r}/t_k\to0$. In this situation the angle $\phi$ varies so fast that on timescales governing the escape dynamics of the particle the directed motion resulting from the drift term $F\cos\phi\,\diffd t$ in \refeq{Langevin1} can safely be neglected. The influence of the particle's propulsion on its translational dynamics reduces then to an enhancement in diffusivity \cite{TenHagen2011,Ao2015}. Consequently, we again can model the escape dynamics via an effective passive particle dynamics, assuming now, however, an effective diffusion constant.

In order to obtain these sought corrections to the particle's diffusivity due to an active propulsion, we use the homogenization mapping procedure detailed in Ref.\ \cite{Kalinay2014} to project the two-dimensional phase space of the Fokker-Planck dynamics (\ref{FPE1Dscaled}) onto a one-dimensional phase space differential equation w.r.t.\ the position coordinate $\xi$. That is, we are looking for an equation for the marginal probability density function
\begin{equation}
\label{fwdmapping}
\mathcal{P}(\xi,\tau):=\int\limits_0^{2\pi}\diffd\phi\,P(\xi,\phi,\tau)\;,
\end{equation}
where the latter reduction of variables is assumed to be reversible by means of the ``backward mapping'' operator $\hat{\omega}(\xi,\phi)$,
\begin{equation}
\label{backward}
P(\xi,\phi,\tau)=\hat{\omega}(\xi,\phi)\frac{\mathcal{P}(\xi,\tau)}{2\pi}\;.
\end{equation}
Here, $\mathcal{P}(\xi,\tau)/(2\pi)$ is the density $P(\xi,\phi,\tau)$ for infinitely fast relaxation in $\phi$-direction, i.e., for infinitely fast particle rotation ($t_\mathrm{r}/t_k=0$). In this case the propulsion force cannot contribute to the translational dynamics anymore, implying that the active particle dynamics renders into a passive one and $P(\xi,\phi,\tau)$ becomes independent of $\phi$. If now $t_\mathrm{r}/t_k=4kR^2/(3\kb T)=:\varepsilon$ is very small, the difference between $P(\xi,\phi,\tau)$ and $\mathcal{P}(\xi,\tau)/\allowbreak(2\pi)$ must likewise be very small. Thus, $\hat{\omega}(\xi,\phi)$ can be expanded in $\varepsilon$ around $\varepsilon=0$, yielding
\begin{equation}
\label{bwdmapping}
P(\xi,\phi,\tau)=\sum_{n=0}^\infty\varepsilon^n\hat{\omega}_n(\xi,\phi)\frac{\mathcal{P}(\xi,\tau)}{2\pi}\;,
\end{equation}
where $\hat{\omega}_0(\xi,\phi)=1$. If we next apply \refeq{fwdmapping} and \refeq{bwdmapping}, respectively, to \refeq{FPE1Dscaled} (where for the sake of convenience we have used the substitutions $\alpha:=\kb T/(6U_0)$ and $\beta:=F/(kx_\mathrm{max})$), we obtain two equations for $\mathcal{P}(\xi,\tau)$, reading
\begin{alignat}{2}
\label{mapeq1}
\frac{\partial\mathcal{P}(\xi,\tau)}{\partial\tau}=&\left(\alpha\frac{\partial^2}{\partial\xi^2}+\frac{\partial}{\partial\xi}\xi(1-\xi)\right)\mathcal{P}(\xi,\tau)\nonumber\\
&-\sum_{n=0}^\infty\varepsilon^n\beta\frac{\partial}{\partial\xi}\piint\diffd\phi\,\cos\phi\,\hat{\omega}_n(\xi,\phi)\frac{\mathcal{P}(\xi,\tau)}{2\pi}
\end{alignat}
and
\begin{alignat}{2}
\label{mapeq2}
&\sum_{n=0}^\infty\varepsilon^n\hat{\omega}_n(\xi,\phi)\frac{\partial}{\partial\tau}\frac{\mathcal{P}(\xi,\tau)}{2\pi}=\nonumber\\
&\sum_{n=0}^\infty\varepsilon^n\left[\alpha\frac{\partial^2}{\partial\xi^2}+\frac{1}{\varepsilon}\frac{\partial^2}{\partial\phi^2}+\frac{\partial}{\partial \xi}\left[\xi(1-\xi)-\beta\cos\phi\right]\right]\nonumber\\
&\quad\ \,\times\hat{\omega}_n(\xi,\phi)\frac{\mathcal{P}(\xi,\tau)}{2\pi}\;.
\end{alignat}
Inserting \refeq{mapeq1} into \refeq{mapeq2} and grouping by powers of $\varepsilon$ in turn yields the operator recurrence relation for the $\hat{\omega}_n$,
\begin{alignat}{2}
\partial_\phi^2&\hat{\omega}_{n+1}(\xi,\phi)=\nonumber\\
&\left[\hat{\omega}_n(\xi,\phi),\left(\alpha\partial_{\xi}^2+\partial_{\xi}\xi(1-\xi)\right)\right]+\beta\cos\phi\,\partial_{\xi}\hat{\omega}_n(\xi,\phi)\nonumber\\
&-\frac{\beta}{2\pi}\sum_{m=0}^n\hat{\omega}_{n-m}(\xi,\phi)\partial_{\xi}\piint\diffd\phi\,\cos\phi\,\hat{\omega}_{m}(\xi,\phi)
\end{alignat}
(acting on the probability distribution $\mathcal{P}(\xi,\tau)$), where $[\bullet,\bullet]$ denotes the commutator of the corresponding two operators.
Using the initial condition that $\hat{\omega}_0(\xi,\phi)=1$, the periodicity condition $\hat{\omega}_{n}(\xi,0)=\hat{\omega}_{n}(\xi,2\pi)$ and the normalization condition $\piint\diffd\phi\,\hat{\omega}_{n}(\xi,\phi)=2\pi\delta_{n,0}$, we iteratively can solve for the sought $\hat{\omega}_n$ up to arbitrarily high order.

Although in the considered limit of fast particle rotation, i.e., for $\varepsilon\to0$, it is sufficient to consider only terms of order $\varepsilon$, an improved result possessing a wider range of validity can be obtained if we collect all terms holding the same structure as the ones of $\mathcal{O}(\varepsilon)$, yielding the compact result
\begin{equation}
\label{summe}
\hat{\omega}(\xi,\phi)=1+\sum\limits_{n=1}^{\infty}(-1)^n\varepsilon^n\beta\cos\phi\,\partial_{\xi}=1-\frac{\beta\varepsilon}{1+\varepsilon}\cos\phi\,\partial_{\xi}\;.
\end{equation}
Thus, upon integrating over the angle $\phi$ in Eq.\ (\ref{FPE1Dscaled}) and subsequently inserting Eq.\ (\ref{backward}) with Eq.\ (\ref{summe}) into it, the projected differential equation describing the temporal evolution of the marginal probability density $\mathcal{P}(\xi,\tau)$ reads explicitly:
\begin{alignat}{2}
\frac{\partial\mathcal{P}(\xi,\tau)}{\partial\tau}=&\left[\frac{\kb T}{6U_0}\left(1+\frac{2F^2R^2}{\kb T\left(4kR^2+3\kb T\right)}\right)\frac{\partial^2}{\partial\xi^2}\right.\nonumber\\
&\ \ +\left.\frac{\partial}{\partial\xi}\xi(1-\xi)\right]\mathcal{P}(\xi,\tau)\;.
\end{alignat}
As expected, the latter equation is equivalent to the Fokker-Planck dynamics of a passive particle, where the diffusivity becomes enhanced due to the presence of active propulsion at work. This enhancement in diffusivity may formally also be accomplished by introducing an effective temperature \cite{Szamel2014}, i.e.,
\begin{equation}
\label{Teff}
T_\mathrm{eff}=T\left(1+\frac{2F^2R^2}{\kb T\left(4kR^2+3\kb T\right)}\right)\;.
\end{equation}
However, it has to be pointed out that effective diffusion constants or effective temperatures are only appropriate to describe the dynamics of a SPP if the particle is in the diffusive regime, since in the non-diffusive regime it has in general a considerably non-Gaussian property \cite{Zheng2013}.
We also remark that the above-noted effective temperature of a SPP in the potential $\xi^2/2-\xi^3/3$ coincides with its effective temperature in the harmonic potential $\xi^2/2$ (where the diffusive approximation preserves the first two moments of the particle's position, which can be analytically calculated from the Langevin formalism \cite{TenHagen2011}), indicating that at the saddle points of the external potential the active propulsion contributes dominantly to the particle's dynamics. Vice versa, the steeper the slope of the potential, the less does the propulsion influence the particle's position. This arises from the fact that for steep potential slopes the propulsion force becomes negligible compared to the gradient of the external potential.

\begin{figure}[t]
\begin{center}\includegraphics[width=0.9\columnwidth]{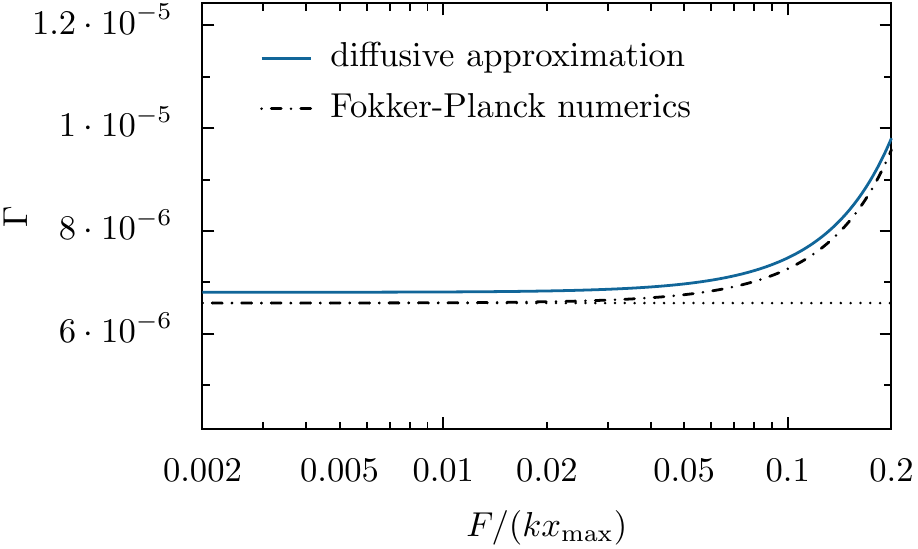}\end{center}%
\caption{\label{F5}(color online) Escape rate of a SPP as a function of the propulsion strength, where the parameters were chosen according to $k=10^{-6}\,\mathrm{N}\mathrm{m}^{-1}$, $T=300\,\mathrm{K}$, $x_\text{max}=0.5\,\mu\mathrm{m}$, and $R=10\,\mathrm{nm}$. The dotted black line indicates again the exact escape rate of a passive Brownian particle. It can be seen that the results obtained under the diffusive approximation coincide well with the exact numerical ones. (The larger divergence noted at first sight compared to Fig.\ \ref{F4} is solely a consequence of the $y$-axis' scaling. Actually, the residual error (i.e., the error for $F=0$) resulting from the approximations underlying the general flux-over-population method is about $4\%$ for the present choice of $U(x)$ and $T$. This discrepancy also appears in Fig.\ \ref{F4}, however there the $y$-axis' scaling is so high that it is hardly noticeable.)}
\end{figure}

Returning to the original objective of studying the escape dynamics, the corresponding escape rate can again be calculated analytically using the flux-over-population method, yielding
\begin{equation}
\label{EscapeRateDiffusive}
\mathrm{\Gamma}=\frac{1}{2\pi}\exp\left[-\frac{U_0}{\kb T}\left(1+\frac{2F^2R^2}{\kb T\left(4kR^2+3\kb T\right)}\right)^{-1}\right]\;.
\end{equation}
Because $U_0/(\kb T)$ denotes the ratio between the potential barrier height and the thermal energy, in the diffusive approximation the escape rate $\mathrm{\Gamma}$ follows a modified Arrhenius' law, where the actual temperature $T$ is replaced by an effective one, defined by \refeq{Teff}.
A graphical comparison between $\mathrm{\Gamma}$ calculated by means of \refeq{EscapeRateDiffusive} and exact numerical results is depicted in \reffig{F5}. The diffusive approximation indeed succeeds in describing the dependence of the particle's escape rate on its propulsion strength for fast particle rotation. Note that the value for $R$ used in \reffig{F5} implies a ratio $t_\mathrm{r}/t_k\approx0.032$.

\begin{figure}[t]
\begin{center}\includegraphics[width=0.85\columnwidth]{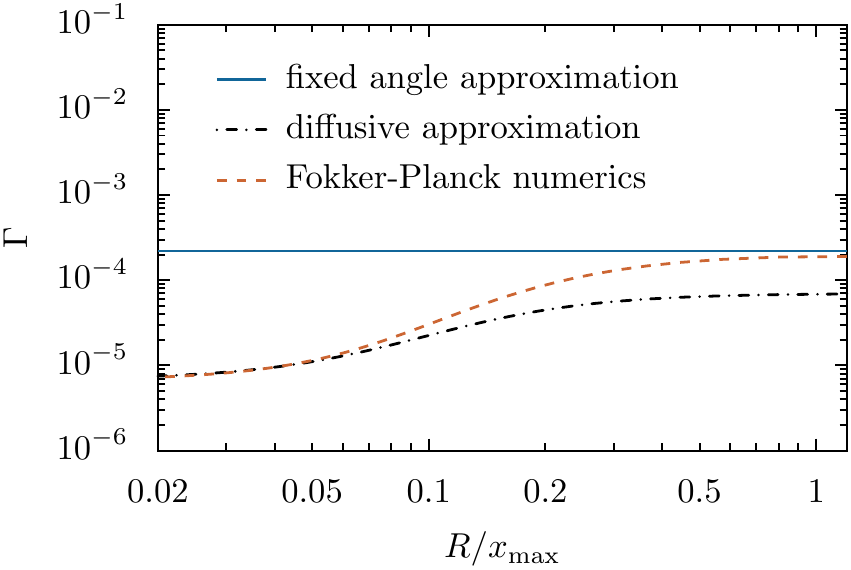}\end{center}%
\caption{\label{F6}(color online) Dependence of the escape rate of a SPP on the particle radius. The choice of parameters was $k=10^{-6}\,\mathrm{N}\mathrm{m}^{-1}$, $T=300\,\mathrm{K}$, $F=5\cdot10^{-14}\,\mathrm{N}$, and $x_\mathrm{max}=0.5\,\mu\mathrm{m}$. One can clearly detect the transition between the ranges of validity of the two analytic approximations.}
\end{figure}

To illustrate the dependence of the escape rate $\mathrm{\Gamma}$ on the particle radius and therefore to determine the range of validity of the diffusive and, as well, the fixed angle approximation, in \reffig{F6} we also depict the escape rate as a function of $R$. Expectedly, the diffusive approximation yields very good results for small particle radii; with increasing $R$, however, the approximation starts to fail. The deviations stem from the fact that the ballistic properties of the particle's propulsion become increasingly relevant. Conversely, for larger radii $R$ the fixed angle approximation, which always provides an upper bound for $\mathrm{\Gamma}$, takes over to describe the actual result for the escape rate.

\section{Summary}
\label{SecSum}

In this work we have investigated the escape dynamics of a self-propelled particle dwelling a metastable potential landscape. In doing so, both numerical and analytical methods have been applied. For fast and slow particle rotation, we were able to derive tractable approximate analytic expressions for the escape rate. Two main parameters were identified to predominantly govern the escape dynamics for given potential and temperature, the strength $F$ of the particle's propulsion force and the particle radius $R$: While $F$ determines how strong the active propulsion may maximally contribute to the dynamics of the particle's position, $R$ governs the ratio between the rotational diffusion time and the relaxation time due to the restoring force of the potential, ruling how much the active propulsion can actually contribute to the displacement of the particle position. For large $R$ the particle rotates rather slowly compared to the timescale of its translational dynamics in the well, why in this instance the fixed angle approximation yields good results. One must however pay attention that the rotational diffusion time may not become larger than the mean escape time of a particle orientated toward the barrier, since otherwise the timescale of a particle's escape would not predominate all other timescales of the particle's dynamics in the well and the escape problem could not be described as a rate process any longer \cite{Hanggi1990,Hanggi1995}.

For a small particle radius $R$ on the other hand, the particle rotates so fast that the active propulsion cannot establish an appreciable drift regime. It solely gives rise to an enhanced diffusivity of the particle, for what reason the diffusive approximation yields very good results.
The case of moderate particle radii however poses a serious problem to analytical approaches, since in this instance both ballistic and diffusive properties of the active propulsion contribute equally to the escape dynamics and therefore neither of them may be neglected.

Finally we remark that because for $F\to0$ the active particle turns into a passive one, the escape rates obtained under the fixed angle approximation and under the diffusive approximation concur in the case of a vanishing driving force, agreeing with the well known, overdamped Kramers rate,
\begin{equation}
\mathrm{\Gamma}=\frac{1}{2\pi}\exp\left(-\frac{U_0}{\kb T}\right)\;,
\end{equation}
of a passive Brownian particle.
\newline
\newline

{\small
We gratefully acknowledge financial support from the cluster of excellence Nanosystems Initiative Munich (NIM).}

\appendix
\section{Numerical Methods}
\numberwithin{equation}{section}
\label{NumMeth}
\subsection{Monte-Carlo simulations}

For our Monte-Carlo simulations, the set of Langevin equations (\ref{Langevin1}), (\ref{Langevin2}) and (\ref{Langevin3}) was integrated numerically by means of the Euler-Maruyama method \cite{Kloeden1992}, where the random numbers representing Gaussian white noise were generated using the \emph{Mersenne twister} algorithm \cite{Matsumoto1998}. We calculated the particle's mean escape time from the potential well by simulating an ensemble of trajectories starting from $\xi(0)=0$ with random initial conditions $\phi(0)\in[0,2\pi]$: once a defined point slightly beyond the top of the barrier is reached by the $i$-th realization, the associated escape time $\tau_{\mathrm{exit},i}$ is detected. The particle's escape rate then follows from the relation
\begin{equation}
\mathrm{\Gamma}=\langle \tau_\mathrm{exit}\rangle^{-1}=\left(\frac{1}{N}\sum_{i=1}^N \tau_{\mathrm{exit},i}\right)^{-1}\;,
\end{equation}
where $N$ denotes the ensemble size.

\subsection{Fokker-Planck formalism}

The escape rate of a self-propelled particle can be calculated as well numerically within the Fokker-Planck formalism. Rewriting Eq.\ (\ref{FPE1Dscaled}) as a continuity equation yields
\begin{equation}
\partial_\tau P(\xi,\phi,\tau)=-\partial_\xi J_\xi(\xi,\phi,\tau)-\partial_\phi J_\phi(\xi,\phi,\tau)\;,
\end{equation}
with the components of the probability current $\mathbf{J}$ being given by
{\small
\begin{equation}
J_\xi(\xi,\phi,\tau)=\left(-\frac{\kb T}{6U_0}\partial_\xi-\xi(1-\xi)+\frac{F}{kx_\mathrm{max}}\cos\phi\right)P(\xi,\phi,\tau)
\end{equation}}%
and
{\small
\begin{equation}
J_\phi(\xi,\phi,\tau)=-\frac{3\kb T}{4kR^2}\partial_\phi P(\xi,\phi,\tau)\;.
\end{equation}}%
Because the particle's escape rate $\mathrm{\Gamma}$ is determined by the probability current in $\xi$-direction on top of the barrier, Eq.\ (\ref{FPE1Dscaled}) was numerically integrated upon combining the method of lines \cite{Schiesser1991} with a second-order back\-ward-dif\-fer\-ence scheme \cite{Hairer1996}. In order to allow for the existence of a stationary solution in the considered metastable potential, the boundary condition
\begin{equation}
J_\xi(\xi_\mathrm{l},\phi,\tau)=J_\xi(\xi_\mathrm{r},\phi,\tau)
\label{NumFOPBC}
\end{equation}
was introduced, where $\xi_\mathrm{l}$ is located leftward of the potential well and $\xi_\mathrm{r}$ to the right of the barrier. The latter periodic boundary condition implies that the probability flowing out over the barrier is ``re-injected'' at $\xi_\mathrm{l}$. The exact position of this injection point however has to be chosen carefully, because the above boundary condition not only allows for the particle to exit at $\xi_\mathrm{r}$ and reenter at $\xi_\mathrm{l}$, but also to perform the process in the opposite direction. Thus, $\xi_\mathrm{l}$ must be located sufficiently to the left of the metastable potential well, so that the event of the particle exiting at the left and entering at the right side of the barrier becomes extremely unlikely. In the $\phi$-direction we also imposed periodic boundary conditions, i.e.,
\begin{alignat}{2}
P(\xi,0,\tau)&=P(\xi,2\pi,\tau)\nonumber\\
\partial_\phi P(\xi,\phi,\tau)|_{\phi=0}&=\partial_\phi P(\xi,\phi,\tau)|_{\phi=2\pi}\;,
\end{alignat}
and for the initial condition we again assumed the particle to be located at the bottom of the well with a uniformly distributed starting angle, $P(\xi,\phi,0)=1/(2\pi)\delta(\xi)$.
The sought escape rate $\mathrm{\Gamma}$ can then be obtained by computing the stationary probability current across the barrier and subsequently integrating over all orientation angles:
\begin{equation}
\mathrm{\Gamma}=\piint\diffd\phi\,\lim\limits_{\tau\to\infty}J_\xi(1,\phi,\tau)\;.
\end{equation}
However, we remark that the technique presented thus far yields correct results only for a sufficiently large ratio between the mean escape time $\mathrm{\Gamma}_{\phi=0}^{-1}$ of particles orientated to the right and the scaled rotational diffusion time $t_\mathrm{r}/t_k$. This is owed to the fact that in Eq.\ (\ref{NumFOPBC}) the particle's orientation is kept when exiting at $\xi_\mathrm{r}$ and re-entering at $\xi_\mathrm{l}$, for what reason in the second iteration its starting angle is not uniformly distributed anymore (particles that have managed to cross the barrier and exit at $\xi_\mathrm{r}$ are more probably orientated to the right). This fact does indeed not pose a problem as long as the particle's sojourn in the potential well is considerably longer than its rotational diffusion time; then, the particle can thermalize inside the well and the memory of the insertion angle gets lost. For rotational diffusion times larger than the mean escape time related to a fixed orientation of $\phi=0$, though, the particle might pass through several iteration cycles without significantly changing its orientation, resulting in an overestimated escape rate.

In order to cover also the case $t_\mathrm{r}/t_k>\mathrm{\Gamma}_{\phi=0}^{-1}$, we introduced an artificial rotational ``thermalization'' which the particle has to undergo when it re-enters the well. The latter can conveniently be modeled by a spatially dependent rotational diffusion coefficient that is strongly increased within a small domain near $\xi_\mathrm{l}$. Thus, we substituted
\begin{equation}
\frac{\partial^2}{\partial\phi^2}P(\xi,\phi,\tau)\to\frac{\partial^2}{\partial\phi^2}\left[1+\frac{1}{\sigma}\exp\left(\frac{(\xi-\xi_\mathrm{l})^2}{2\sigma^2}\right)\right]P(\xi,\phi,\tau)
\end{equation}
in Eq.\ (\ref{FPE1Dscaled}), where $\sigma$ has to be chosen very small. Now, all particles entering at $\xi_\mathrm{l}$ and moving toward the bottom of the potential well must pass an orientation-equalizing area, consequently the artifact resulting from a non-uni\-form\-ly distributed insertion angle vanishes. As the region where the rotational diffusion coefficient deviates from its true value is very small and notably is located rather apart from the potential well, this strategy does not distort the results markedly.

\bibliographystyle{epj}
\bibliography{library}

\end{document}